\documentclass[aps,pra,twocolumn,floatfix]{revtex4-2}
\usepackage{amsmath,epsfig,mathrsfs,graphicx,amssymb,color,float}
\usepackage[T1]{fontenc}
\usepackage{bbm}
\usepackage{t1enc}
\usepackage{hyperref}
\usepackage{array}
\usepackage{booktabs}
\newtheorem{thm}{Theorem}

\def\be#1\ee{\begin{equation}#1\end{equation}}
\newcommand{\ba}{\begin{eqnarray} }
\newcommand{\ea}{\end{eqnarray} }

\usepackage[cp1250]{inputenc}   
\usepackage{amsmath}            
\usepackage{amsfonts}           
\usepackage{amssymb}            
\usepackage{enumerate,graphicx}
\usepackage{array}
\usepackage{booktabs}

\usepackage{physics}

\usepackage{amsmath}            
\usepackage{amsfonts}           
\usepackage{amssymb}            
\usepackage{enumerate,graphicx}
\usepackage{array}
\usepackage{booktabs}
\usepackage{yquant}
\usetikzlibrary{fit, quotes}
\usetikzlibrary{snakes,arrows,shapes}
\useyquantlanguage{groups}
\usepackage{physics}
\usepackage{todonotes}

\def\mb{\begin{pmatrix}}
\def\me{\end{pmatrix}}
\def\be#1\ee{\begin{equation}#1\end{equation}}

\def\d{\textrm{d}}

\def\mb{\begin{pmatrix}}
\def\me{\end{pmatrix}}
\def\be#1\ee{\begin{equation}#1\end{equation}}

\def\d{\textrm{d}}

\newcommand{\AB}[1]{{{  {#1}}}}
\begin{document}

\title{Identical, independent quantum weak measurements violate objective realism}

	\author{Tomasz Rybotycki$^{1,2,3}$}
	\author{Tomasz Bia{\l}ecki$^{4,5}$}
	\author{Josep Batle$^{6,7}$}
	\author{Bart{\l}omiej Zglinicki$^4$}
	\author{Adam Szereszewski$^4$}
	\author{Wolfgang Belzig$^8$}
	\author{Adam Bednorz$^4$}
	\email{abednorz@fuw.edu.pl}
	
	\affiliation{$^1$Systems Research Institute, Polish Academy of Sciences, ul. Newelska 6,
		01-447 Warsaw, Poland}
	\affiliation{$^2$Nicolaus Copernicus Astronomical Center, Polish Academy of Sciences,
		ul. Bartycka 18, 00-716 Warsaw, Poland
	}
	\affiliation{$^3$Center of Excellence in Artificial Intelligence, AGH University,
		al. Mickiewicza 30, 30-059 Cracow, Poland
	}
	\affiliation{$^4$Faculty of Physics, University of Warsaw, ul. Pasteura 5, PL02-093
		Warsaw, Poland}
	\affiliation{$^5$Faculty of Physics and Applied Informatics, University of Lodz,
		ul. Pomorska 149/153, PL90-236 Lodz, Poland}
	\affiliation{$^6$Departament de Física, Universitat de les Illes Balears E-07122 Palma de Mallorca, Balearic Islands, Spain}
	\affiliation{$^7$CRISP -- Centre de Recerca Independent de sa Pobla, 07420 sa Pobla,
		Balearic Islands, Spain}
	\affiliation{$^8$Fachbereich Physik, Universit{\"a}t Konstanz, D-78457 Konstanz, Germany}

	\begin{abstract}
	We demonstrate violation of objective realism in quantum world using unconstrained weak measurements. Instead of 
	limited Leggett-Garg approach with artificial bounds on the observed values, we assume two identical and indepenent weak detectors
	and final conditioning. The experimental verification has been performed on public quantum computers, IBM and IonQ. Thanks to sufficiently large statistics,
	the violation is observed at the level of 10 standard deviations. The tests confirmed also high quality of parametric two-qubit gates offered by
	main quantum hardware providers.
	
	\end{abstract}
\maketitle

\emph{Introduction.}
Quantum experiments cannot be explained by an underlying local classical theory, \cite{epr,bell,chsh}. 
Theoretical description does not give a consistent picture of objective reality~\cite{mer}, especially in the case of incompatible (non-commuting) observables. An example is Wigner function of position and momentum~\cite{wigner}, which --- in contrast to classical probability --- can be negative.
The origin of this conflict is the lack of natural noninvasive measurements,
which would allow to observe the system during its evolution. Any information gain results in some disturbance~\cite{busch}.
Moreover, continuous strong, projective measurements usually lead to freezing the dynamics (quantum Zeno effect \cite{zeno}).

A direct failure of quantum realism is observed by Leggett-Garg (LG) inequality~\cite{lega,emary}
which checks if the measurements disturb the system. The violation of LG inequality  by strong, projective measurements \cite{lginv1,lginv2},
 obviously triggering a quantum state collapse~\cite{neumann} is not surprising. However, it is also violated by weak measurements
~\cite{lgw1,jor1,jor2,semi3,palacios-laloy:10,gog,semi1,semi2,semi4,lgw2,semi1,semi2}.
The weak measurement~\cite{aav} is performed by a detector only slightly interacting with the probed system. In the limit of vanishing 
interaction, there is no disturbance. The trade-off between information gain and the disturbance allows one to find an interaction scale at which
the measurement strength $\lambda$ results in the disturbance $\sim\lambda^2$, making weak measurements the closest counterparts of classical
ideal measurements~\cite{bednorz:10,abn}. Apart from the LG inequality, weak measurements also violate time-order symmetry~\cite{bfb},

The LG inequality needs the variables to be limited by $\pm 1$. This assumption is natural for projective measurements of a 
dichotomic observable, but a weak measurement produces a noisy detector pointer with typically unbounded outcomes, and with direct averages and correlations $\ll 1$.
To obtain a violation one requires an additional rescaling by an independently calibrated measurement strength \cite{palacios-laloy:10,gog,semi1,piacentini,semi2,semi4}.

Here we present an inequality without this limitation. Instead, one applies the same weak measurement twice with two identical, independent detectors and uses their cross-correlations (together with final conditioning). Our inequality does not assume the measurement strength but requires the correlation from the two measurements, which is experimentally demanding, as the necessary statistics for a sufficient confidence level scale with the inverse square of the strength.
Nevertheless, we were able to confirm the violation with public quantum computers, IBM and IonQ, at a high confidence level,
using natural strength-dependent two-qubit gates, with similar circuits as in the test of the standard LG inequality and time symmetry \cite{epj}. 

\emph{Weak measurement.} A general quantum measurement is represented by an instrument \cite{inst1,inst2,inst3}, 
	a collection of trace-nonincreasing linear completely positive maps $\mathcal K(a)$ where $a$ is the
	measurement outcome~\cite{kraus,povm,peres,nielsen}. Then the state $\rho$ becomes
	\be
	\rho(a)=\mathcal K(a)\rho\label{rab}
	\ee
	with probability $p(a)=\mathrm{Tr}\rho(a)$, and normalized by $\int p(a) \d a=1$ or equivalently $\mathrm{Tr}\mathcal K\rho=1$
	for the quantum channel $\mathcal K\equiv \int \mathcal K(a)\d a$, a Hermitian $\rho\geq 0$, $\mathrm{Tr}\rho=1$.  The disturbance on a generic state $\rho$ 
	can be then described by the state after the measurement
     $\mathcal K\rho$ with the results discarded, so that in general it is a mixed state.

	We shall denote classical and quantum averages $\langle f(a)\rangle\equiv \int \d a\;
	f(a)p(a)$ and $\langle \mathcal F\rangle=\mathrm{Tr}\mathcal F\rho$, respectively.
	Every measurement can be realized by system space extended by an auxiliary detector (meter) \cite{kraus,povm,nielsen}, 
	\be
	\mathcal K(a)\rho=\mathrm{Tr}'M(a)\check{\mathcal U}(\rho P)\label{kra}
	\ee
	where $P$ is the initial state of the detector (meter), $\mathrm{Tr}P=1$, $M(a)$ is the measurement operator 
	(effect) for the outcome $a$ in the meter 
	space, $\int M(a)\d a=1$, $\check{\mathcal U}A=UAU^\dag$ is a unitary (adjoint) map for unitary $U$, 
	and the trace $\mathrm{Tr}'$ is partial, over the meter space. 
	Here $P,M(a)\geq 0$ are Hermitian and we drop the tensor sign $\otimes$ by putting the product in brackets $()$.
	The use of meters allows to measure the same system sequentially many times.
	A final direct measurement can be realized by $\mathcal K(a)\rho=\mathrm{Tr}M(a)\rho$  defined within the system space ($M=1$ if the final measurement is ignored).
	
	A completely noninvasive measurement occurs only when $\check{\mathcal U}=1$ resulting in scalar $\mathcal K(a)=\mathrm{Tr}M(a)P\equiv k(a)$ containing only 
the internal noise of the meter. Such event is commonly referred to as no information gain without disturbance~\cite{busch,epj}. By shifting $a$ we can assume that this noise is centered at $\langle a\rangle=0$.
	We can still consider $\mathcal K(a)\simeq k(a)$ which defines a class of weak measurements.
	 We further assume that the information obtained from the
	system is encoded in the average $\langle a\rangle$ for $U\neq 1$. 
	We also generalize the average based on (\ref{kra}) to a class of averaging maps
	\be
	\bar{\mathcal K}=\mathrm{Tr}'\bar{M}\check{\mathcal U}(\rho \bar{P})\label{axx}
	\ee
	where $\bar{M}=\int aM(a)\d a$ and $\bar{P}=\sum_j \bar{p}_jP_j$ with many meter  state preparations $P_j$. The single measurement effect case means $\bar{p}_1=1$ 
	but a very useful choice is a \emph{contrast measurement} $\sum_j\bar{p}_j=0$, giving $\mathrm{Tr}\bar{P}=0$. \AB{Note that $\bar{p}_j$ are real but not always postive and the goal of this combination is to remove possible bias of detector initialization}. Adjusting the shift of $a$
	we set $\bar{\mathcal K}=0$ for $\mathcal U=1$, which means no bias. We have the connection with correlations \AB{measurable by meter's $M(a)$},
	\be
	\langle\cdots \bar{\mathcal K}\cdots\rangle=\sum_{j}\bar{p}_j\langle \cdots a\cdots \rangle_j\equiv \langle \cdots a\cdots \rangle.
	\label{cont}
	\ee
	This notation unifies the case of a single and many preparations.
	
	The weak measurement can be defined by scalable evolution $U^\lambda=\exp\lambda H/i$ for Hermitian $H$,
	equivalent to $\check{\mathcal U}^\lambda=\exp\lambda \tilde{\mathcal H}$, 
	with $\tilde{\mathcal A}B=i[B,A]$, and commutator notation $[A,B]\equiv AB-BA$. Here
	$\lambda\to 0$ is a combination of time and interaction strength. Then
	the evolution expansion in $\lambda$ in the lowest order reads
	\be
	\check{\mathcal U}\simeq 1+\lambda \tilde{\mathcal H}.\label{hhh}
	\ee 
	Then, for small $\lambda$, one is left with the generic form \cite{bednorz:10,abn,bfb,povmpp,buelte:18},
	\be
	\bar{\mathcal K}\simeq \lambda(\hat{\mathcal A}+\tilde{\mathcal A'})\label{avv}
	\ee
	denoting $\hat{\mathcal A}B=\{A,B\}/2$ with the anticommutator $\{F,G\}\equiv FG+GF$. 
	The above remains true even if one adds some randomness of $\check{\mathcal U}$.

	\AB{The maps $\hat{\mathcal A}$ and $\tilde{\mathcal A'}$ depend on Hermitian operators $A$ and $A'$, respectively, which depend on $H$ and } can be extracted from 
	\be
	A'\mathrm{Tr}'\{\bar{M},\bar{P}\}H/2,\quad A= i \mathrm{Tr}' [\bar{M},\bar{P}]H.\label{aap}
	\ee
	The corrections to (\ref{avv}) are generally $\sim\lambda^3$
	because of unavoidable disturbance $\mathcal K\sim \lambda^2$ for $\sim\lambda$ terms calibrated to zero by $\mathrm{Tr}'HP=0$. Their magnitude depends on the concrete detection scheme \cite{epj}. 
	Since the single weak measurement gives only information about $A$, because
	\be
	\langle a\rangle\simeq \lambda\langle A\rangle, \label{info}
	\ee
	we shall call the case $A'=0$ \emph{informative} weak measurement.	
	
	The nonzero $A'$ affects the next measurement or
	postselection, corresponding to the influence of the detector on the weakly measured system.
	This can also occur in classical indirect measurements \cite{bfb} and special measurement schemes \cite{dj12,povmpp,buelte:18}. If one applies an external parametric influence $U'=\exp gA'/i$ on the system, then the systems' response reads
	\be
	\left.\frac{\d\langle \cdots\rangle}{\d g}\right|_{g=0}=\langle \cdots \tilde{\mathcal A}'\cdots\rangle,
	\ee
	so $A=0$ case can be called a \emph{responsive} measurement.
	Responsive and informative measurements can be verified experimentally by the following theorem.
	
	\begin{thm}
	Let two systems $A$ and $B$ be prepared in states $\bar{P}_{A/B}$, measured by $\bar{M}_{A/B}$.
	One can apply a weak measurement for an arbitrary $H$,
	by $\check{\mathcal U}^\lambda=\exp\lambda \tilde{\mathcal H}$.
	The averages and the correlation read
	\begin{align}
	&\langle a\rangle_\lambda=\mathrm{Tr}\bar{M}_A\check{\mathcal U}^\lambda(\bar{P}_A\bar{P}_B),\nonumber\\
	&\langle b\rangle_\lambda=\mathrm{Tr}\bar{M}_B\check{\mathcal U}^\lambda(\bar{P}_A\bar{P}_B),\nonumber\\
	&\langle ab\rangle_\lambda=\mathrm{Tr}\bar{M}_A\bar{M}_B\check{\mathcal U}^\lambda(\bar{P}_A\bar{P}_B).
	\end{align}
	Let these measurements be unbiased, i.e. $\langle a\rangle_0=\langle b\rangle_0=0$.
	Suppose $\lim_{\lambda\to 0}\langle b\rangle_\lambda/\lambda\neq 0$ for some $H$. If  
	$\lim_{\lambda\to 0}\langle a\rangle_\lambda/\lambda\neq 0$ for every $H$, then $A$ makes a weak responsive measurement. If
	$\lim_{\lambda\to 0}\langle ab\rangle_\lambda/\lambda= 0$,
	for an arbitrary $H$, then $A$ makes a weak informative measurement. 
	\end{thm}
	The roles of $A$ and $B$ can be swapped and the resulting type of measurement holds also for any external system. 
	The proof is in Supplemental Material \ref{app0}.	
	The direct construction of informative measurement is possible for two-level systems  by another theorem:
	
	\begin{thm}
	Let the detector be a two-level system, and we have the probabilities for $3$ generic initial states $P_j$, $j=1,2,3$ of the detector
	and a single generic 4-outcome measurement $M_k$, $k=1,2,3,4$, for some weak measurement protocol, with zero and sufficiently small but unknown 
	strength $\lambda$. From these probabilities, one can find the outcome reassignments $k\to\bar{m}_k$, and coefficients $\bar{p}_j$, unique up to a constant 
	factor such that the weak measurement constructed from $\bar{M}$ and $\bar{P}$ is informative.
	\label{tt1}
	\end{thm}
	Here generic means that the Haar measure of the failing cases is zero.
	The proof and its spin-offs are in Supplementary Material \ref{appa}. However, direct application of the theorem
	is  more resource demanding than a partial trust in $\bar{P}$, $\bar{M}$, and $\check{\mathcal U}$.
	
	\begin{figure}
		 \includegraphics[scale=.4]{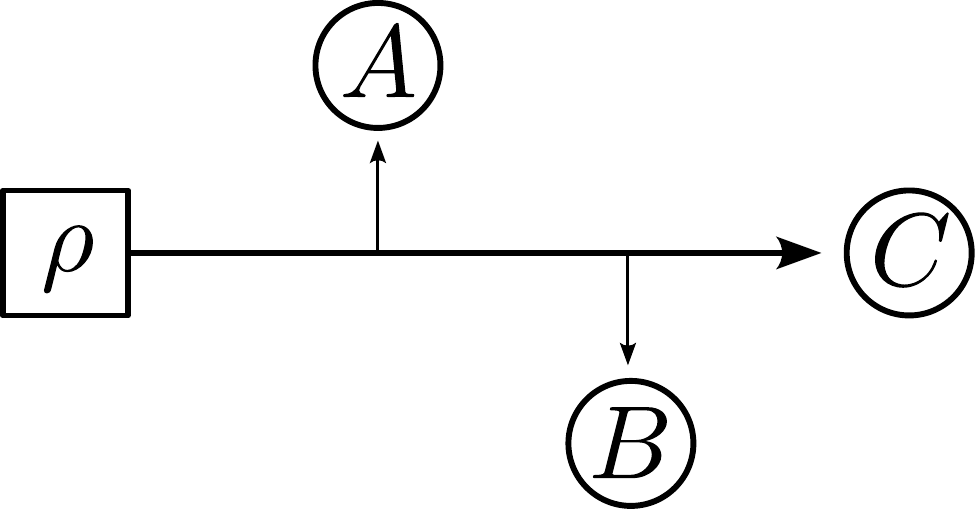}\\
		\medskip
		 \includegraphics[scale=.4]{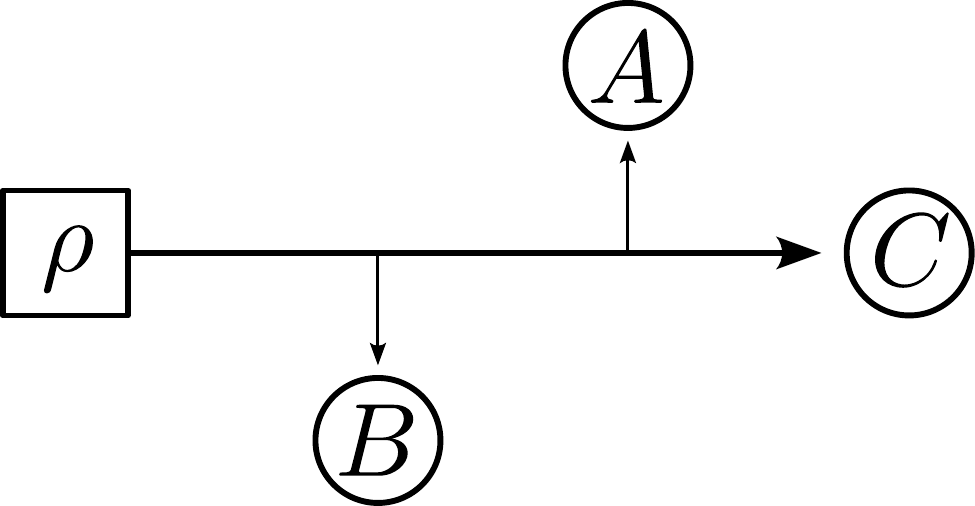}
		\caption{The detection scheme to test objective realism. Two weak detectors of $A$ and $B$ measure the state $\rho$ before the
			final measurement $C$. The time order of the measurement is from the left to the
			right: $A\to B\to C$ in the upper and $B\to A\to C$ in the lower case.}\label{lgw}
	\end{figure}

\emph{Objective reality.} We consider a protocol being a sequence of three measurements, two noninvasive ones, retrieving information of the same objective value, and the final the condition, see Fig. \ref{lgw}, with the corresponding outcomes, $a$, $b$, $c$, of probability $p(a,b,c)$. The condition is dichotomic, $c=1$ if met and $0$ otherwise. The conditional probability reads $p(a,b|c)=p(a,b,c=1)/p(c=1)$ and conditional average/correlation is
	$\langle ab|c\rangle=\langle abc\rangle/p(c=1)$. 
 	
	The following inequality
	\begin{equation}
		\frac{\langle a+b|c\rangle^2}{4\langle ab|c\rangle}
		\leq 1.
		\label{lgg2}
	\end{equation}
	is satisfied classically for an observable $z$ shifting the measurement outcomes
       $a=z+\xi_a$, $b=z+\xi_b$ with uncorrelated unbiased random noise variables $\xi_{a,b}$,  as $\langle a|c\rangle=\langle b|c\rangle=\langle z|c\rangle$
and $\langle ab|c\rangle=\langle z^2|c\rangle$, while $\langle z|c\rangle^2\leq\langle z^2|c\rangle$ equivalent to nonnegative conditional variance.
	This is the expectation from the noisy independent and identical measurements if $z$ is objectively real.
	It remains true for combinations of detector states $j$, with outcomes $a=z_j+\xi_{aj}$, the total observable $z=\sum_j\bar{p}_j z_j$, averages 
	as in (\ref{cont}), and similarly for $b$.
	We will show that it can be violated for quantum identical noninvasive and independent informative measurements. 
	The direct probability of the readout is $p(a,b,c=1)=\langle C\mathcal K_B(b)\mathcal K_A(a)\rangle$ and
	the quantum average reads
	$\langle abc\rangle=\langle C\bar{\mathcal K}_B\bar{\mathcal K}_A\rangle$, in particular $p(c=1)=\langle C\rangle$ if we ignore disturbance of $A$ and $B$.
	Let the measured observables be in the two-level space, $|0\rangle$, $|1\rangle$,
	\begin{align}
	&A=B=Z=|0\rangle\langle 0|-|1\rangle\langle 1|,\nonumber\\
	&C=|\psi_-\rangle\langle\psi_-|
	\end{align}
	and the initial state initial state $\rho=|\psi_+\rangle\langle\psi_+|$ with
	\begin{equation}
	|\psi_\pm\rangle=\cos\frac{\psi}{2}|1\rangle\pm\sin\frac{\psi}{2}|0\rangle.
	\end{equation}
	We consider a sequence of two informative weak measurements $\bar{\mathcal K}_{A,B}\simeq \lambda\hat{\mathcal Z}$, $\lambda \to 0$, and
	the direct last measurement, $M(1)=C$ while $M(0)=1-C$ for $1\geq C\geq 0$, defining the condition. 
	Such a setup is very similar to the original weak value paradox \cite{aav}, where
	the anomaly appeared for the initial and final states almost orthogonal but in contrast we do not assume any bounds on observables.
	Then we get
	\begin{align}
		&\langle C\rangle=\cos^2 \psi,\;\langle \hat{\mathcal Z}^2\rangle=1,\nonumber\\
		&\langle \hat{\mathcal Z}\rangle=\langle C\hat{\mathcal Z}\rangle=-\cos \psi,\nonumber\\
		&2\langle C\hat{\mathcal Z}^2\rangle=\cos^2 \psi+1\label{lca}
	\end{align}
	which violates (\ref{lgg2}) , identifying $\bar{K}_{A,B}\to\lambda \hat{\mathcal Z}$ in the limit $\lambda \to 0$, as the left hand side reads
$2/(\cos^2 \psi+1)$. The largest violation is $2$ at $\psi=\pi/2$ but it becomes then also indeterminate $0/0$.

	Note that the measurements must be informative, because taking responsive measurements one can violate (\ref{lgg2}) even classically, see Supplemental Material 
	\ref{apcl}. On the other hand, (\ref{lgg2}) can be violated by informative measurements of position and momentum, see Supplemental Material \ref{apw}.

\emph{Implementation on IBM and IonQ.}
	IBM Quantum offers networks of qubits, i.e. local two-level spaces, with the default
	basis $|0\rangle$, $|1\rangle$. The states can be changed by unitary operations (gates).
	The qubit is usually measured projectively in the default basis, as the states
	differ by energy and lose mutual coherence over time. However, all projections are
	feasible, by the use of quantum gates. To realize weak measurement on a qubit, it must
	be entangled with an auxiliary meter qubit by a two-qubit operation/gate.

	For the physical implementation and manipulation of qubits as transmons \cite{transmon}
	and programming of IBM Quantum, see \cite{ibm,qis,gam0,qurev,qisr}. To describe the actual
	implementation of the test, we start from  Pauli matrices $X$, $Y$, $Z$,  traceless, Hermitian, satisfying $X^2=Y^2=Z^2=1$, and $XY=iZ$. 
	The eigenbasis of $Z|\pm\rangle=\pm|\pm\rangle$ gives the normalized states
	$|+\rangle\equiv |0\rangle$, $|-\rangle\equiv |1\rangle$, the natural convention  switching between spin and bit notation in quantum computing.
	
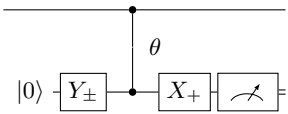
\begin{figure}
		\begin{tikzpicture}[scale=1]
			\begin{yquant*}
				qubit {} q[3];
				discard q[1,2];
				init {$\ket 0$} q[2];
				box {$Y_\pm$}  q[2];
				[shape=yquant-circle, radius=0mm]
				box {} q[1] |  q[0],q[2];
				hspace {-3mm} q[1];
				text {$\theta$} q[1];
				box {$X_+$}   q[2];
				measure q[2];
			\end{yquant*}
		\end{tikzpicture}
		
		\caption{Protocol of weak measurement of $Z$ on the upper qubits with a  $(ZZ)_\theta$ gate by the lower (meter) qubit with the strength of the measurement
			defined by $\sin\theta$. The $(ZZ)_\theta$ gate is depicted as a link between the
			qubits. The sign of the gate $Y_\pm$ is equivalent to switching of the strength $\pm\theta$.}
		\label{weakf}
	\end{figure}

\begin{figure}
		\includegraphics[scale=.8]{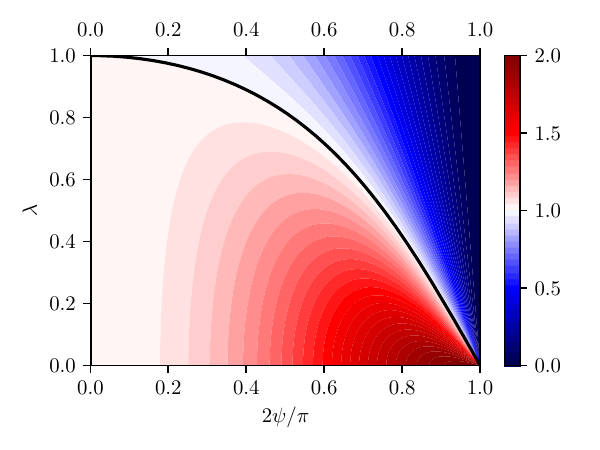}\\

		\caption{The dependence of the ratio
 $\langle a+b|c\rangle^2/4\langle ab|c\rangle$
on the angle $\psi$ between the initial and final state and the measurement strength $\lambda$ which enters by the dependence 
of $p(c=1)=\langle C\rangle_\lambda$
as in (\ref{ccc}), (\ref{axx}), and (\ref{lca}).
The solid line is the threshold $1$. Note that the point $\lambda=0$, $\psi=\pi/2$ is indeterminate, $0/0$.}\label{lgp}
	\end{figure}

	The explicit form of Pauli matrices is given in Supplemental Material \ref{appb}.
	The IBM Quantum devices use transmon qubits and the  native operations (gates) have the form of unitary rotations
	$V_\theta=\exp\theta V/2i$ for Hermitian $V$, and shorthand $V_\pm= V_{\pm \pi/2}$.
	Such systems natively support a single-qubit gate $X_+$ and a virtual phase shift $Z_\theta$ passed to next gates. A sequence of these
	operations allows one to realize an arbitrary unitary $U$. The standard projective
	measurement is implemented by $P_a=|a\rangle\langle a|$ for $a=0,1$. One can
	also see that eigenvalues of $Z=P_0-P_1$ represents the measurement outcomes $z=\pm 1$. The default prepared state is $P_0$  and measurements are $P_{0,1}$
	($0\to +1$, $1\to -1$).
	However, the initialization and measurement can be done along any axis by using a unitary rotation  $\check{\mathcal{U}}_PP_0$  and 
	$\check{\mathcal U}_M^\dag P_{0,1}$. 
	In particular the system's initial state $\rho=\check{\mathcal U}P_0$ and the final 
	strong measurement operator reads $C=\check{\mathcal U}^\dag P_0$ with $U=Y_{\pi/2+\psi}$.	
	To measure weakly, we have to couple the system qubit with an auxiliary meter. We shall
	describe two-qubit gates using the following shorthand notation
	$(AB)|ab\rangle\equiv A|a\rangle B|b\rangle$. The protocol on Heron and IonQ uses native symmetric fractional gates $(ZZ)_\theta$,
	see also Supplemental Material \ref{appb}. The default initial state and measurements can be changed by the unitary rotation.
	The initial detector's state is $Y_\pm|0\rangle=|\pm_x\rangle$, $P_\pm=|\pm_x\rangle\langle \pm_x|$, and
	the measurement is along $Y$ basis by applying $X_+$ before the measurement, $M_\pm=|\pm_y\rangle\langle\pm_y|$ for $|\pm_y\rangle=X_\mp|0\rangle$.
	While IonQ allows to switch the sign of $\theta$, IBM allows only $\theta>0$ and the reversion is implemented by
	the changing sign by $Y_+ \to Y_-$, see Fig. \ref{weakf}.
	We get instruments $\mathcal K_s(z)$ for each choice $s=\pm$ and outcomes $z=\pm 1$ in the form
	\begin{align}
	&2\mathcal K_s(z)=zs\lambda\hat{\mathcal Z}+\mathcal K,\nonumber\\
	&\mathcal K=1+(1-\sqrt{1-\lambda^2})\tilde{\mathcal Z}^2/4.\label{dich}
	\end{align}
	for $\lambda=\sin\theta$.
	This instrument also captures projection at $\lambda=1$ ($\theta=\pi/2$).
	Now $\bar{\mathcal K}=\lambda\hat{\mathcal Z}$ with $\bar{p}_\pm=\pm 1/2$, i.e. $2\bar{P}=P_+-P_-$ in (\ref{axx}).
	Only the average $\langle C\rangle$ gets affected by the finite measurement strength.
	For identical measurements $A$ and $B$ we get the corrected value
	\be
	\langle C\rangle_\lambda=\cos^2 \psi+(\lambda^2\sin^2 \psi) /2.\label{ccc}
	\ee
	This leads to a trade-off between the strength and the angle $\psi$ in (\ref{lgg2}), see Fig. \ref{lgp}.
	 Note that the violation disappears for projective measurements as then $\langle C\rangle_1=\cos^2 \psi+(\sin^2 \psi)/2$
	which gives $2\cos^2\psi/(1+\cos^2 \psi)^2\leq 1$. The measurements are informative relying on the trust in faithful phase shifts $Z_\pm$ and 
	the form of $ZZ$ gate. Imperfect preparation and/or readout and measurement in $Z$ basis and overrotation of $X_+$ gates result
	only in a slight rescaling of $\bar{K}$, see Supplemental Material \ref{appc}.

	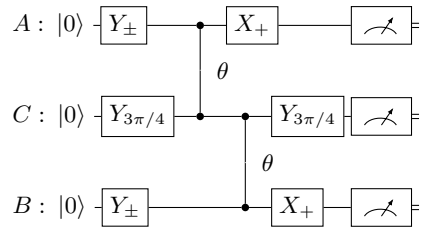
\begin{figure}
		\begin{tikzpicture}[scale=1]
			\begin{yquant*}
				qubit {} q[5];
				discard q[0,1,2,3,4];
				align q[0],q[2],q[4];
				init {$C:\:\ket 0$} q[2];
				init {$A:\:\ket 0$} q[0];
				init {$B:\:\ket 0$} q[4];
				box  {$Y_{3\pi/4}$} q[2];
				box {$Y_\pm$}  q[0];
				[shape=yquant-circle, radius=0mm]
				box {} q[1] |  q[0],q[2];
				hspace {-3mm} q[1];
				text {$\theta$} q[1];
				box {$X_+$}   q[0];				
				box {$Y_\pm$}  q[4];
				[shape=yquant-circle, radius=0mm]
				box {} q[3] |  q[4],q[2];
				hspace {-3mm} q[3];
				text {$\theta$} q[3];
				box {$X_+$}   q[4];
				
				box  {$Y_{3\pi/4}$} q[2];
				align q[0],q[2],q[4];
				measure q[0];
				measure q[2];
				measure q[4];
			\end{yquant*}
		\end{tikzpicture}
		
		\caption{The complete circuit testing objective realism. The system $C$ is measured weakly by meters $A$ and $B$ before the final direct measurement.}
		\label{weaka}
	\end{figure}

	\begin{figure}[ht]
		\includegraphics[scale=.7]{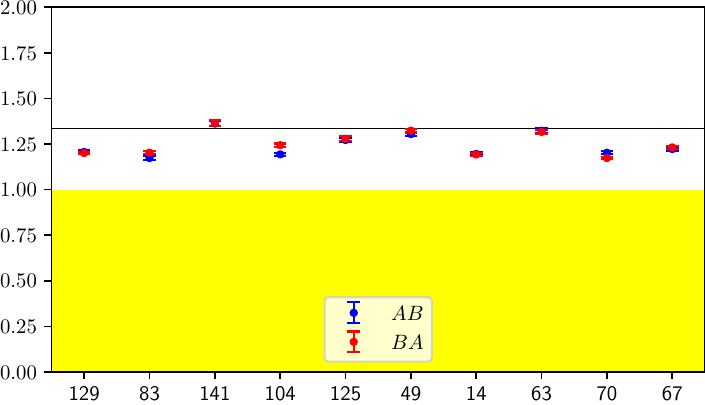}
		\caption{The violation  of (\ref{lgg2}) for fractional $RZZ$ gate on \texttt{ibm\_pittsburgh} for each qubits set 
with the system qubit $C$ index denoted on the horizontal axis.
			The values correspond to the combination of correlations taken from the left hand side of Eq. (\ref{lgg2}), while $AB$ and $BA$ correspond to the order $A\to B$ and $B\to A$.
			The solid line is the perfect value $4/3$ while the yellow region is classical.
		}
		\label{lgt}
	\end{figure}

\emph{Results.}	
	We have tested 10 sets of 3 qubits on IBM Heron device  \texttt{ibm\_kingston}, see the detailed data in Supplemental Material \ref{appd}. We ran 80 jobs, each one with 10000 shots, 25
	repetitions of $A\to B\to C$ and $B\to A\to C$ measurement orders for $\theta=\pm 0.1$
	(i.e. $2\times 2$ measurements), realized by the fractional $(ZZ)_\theta$ gate,  and $\psi=\pi/4$, which should ideally give the left hand side of (\ref{lgg2}) 
	$4/3>1$. 
	The actual measurement is the projection on the states
	$|abc\rangle$ where $a$, $b$ are the outcomes of auxiliary meter qubits, corresponding
	to weak measurements $A$ and $B$, respectively, while $C$ is the final projection on
	the system qubit. To eliminate drifts, we have taken the difference of results for
	$\theta=\pm 0.1$ as in (\ref{axx}) with $P_\pm$ prepared by $Y_{\pm}$ in the weak measurement circuit, see Fig. \ref{weakf} and the full circuit in Fig. \ref{lgt}.  It gives a
	total of 8 circuits per measurement scheme, with $25\times 8=200$ circuits in each job,
	randomly shuffled to avoid memory-related effects. 
	
	The results of the test are shown in Fig.
	\ref{lgt}.  To calculate the error, we assumed that shots are identical and
	independent of each other. This allowed us to use the Bernoulli formula in our analysis. For
	weak measurements, $A$ and $B$, the actual product of values $ab$ is
	almost uniform with random values $\pm 1$. This makes the almost identical error of each of correlations $Q$ of the form
	$\langle cba\rangle$.
	We add up all 4   strength combinations to get the final error scaled by the total number of experiments,
	\be
	\sqrt{\langle (Q-\langle Q\rangle)^2\rangle}\simeq  \sqrt{\langle c\rangle-\langle c\rangle^2}/2\sqrt{JSR}
	\ee
	for $J$ jobs $S$ shots and $R$ repetitions.  It is
	much smaller than the observed violations.	
	We neglected the single average errors, $\langle ca\rangle$ and $\langle cb\rangle$, 
	as their contribution is scaled by the factor $\sim\lambda$.		
	The violation is beyond 12 standard deviations.  
	
	We have also tested IonQ, Forte 1, on a single set of 3 qubits. 
	Formally those were $0,1,2$, but ion trap technology makes them all equivalent. In those
	experiment we ran 1 job, 100000 shots and 25 repetitions each. The circuit was 
	formally identical to the one run on IBM Heron. The tested device contains 36
	ions ${}^{171}\mathrm{Yb}^+$, with the drive frequency 12.64GHz between hyperfine
	levels, in a linear Paul trap with expected average 1 and 2--qubit gate
	errors $0.02\%$ and $0.44\%$, respectively. For technical details see the documentation
	\cite{iondoc,avsion,wright_phd,egan_phd,debnath_phd,landsman_phd,figgatt_phd}.
	
	The results, the left hand side of (\ref{lgg2}), are $1.272\pm 0.035$ in the $A\to B$ case, and
	$1.289 \pm 0.035$ in the $B\to A$ case, show violation of the inequality. Due to longer gate times and repetition rates,
	the whole test took about a month, comparing to several hours on IBM.	
	All correlations and averages are presented in Supplemental Material \ref{appd}.
	The data and the scripts we used are available publicly \cite{zen}.
	
	\emph{Conclusions.}
	Construction of an appropriate inequality allowed to demonstrate violation of objective realism in quantum systems, in particular
	public quantum computers. Due to weak but otherwise unconstrained measurements, feasible experimentally, we were able to collect sufficient level
	of confidence with limited resources. The combination of well-tailored theoretical protocols and state-of-the-art quantum technology permits to exhibit
	fragile nonclassical properties of the quantum world.
	
	\emph{Acknowledgements.}
		The results have been created using IBM Quantum. The views expressed are those of the
	authors and do not reflect the official policy or position of the IBM Quantum team.
	TR gratefully acknowledges the funding support by program "{}Excellence
	initiative -- research university"{} for the AGH University in Krakow as well as the
	ARTIQ project: UMO-2021/01/2/ST6/00004 and ARTIQ/0004/2021.We thank 	
	Pozna{\'n} Supercomputing and Networking Center	for the access to IBM Quantum Innovation Center.

\clearpage

\begin{center}
\large\bf Supplemental Material
\end{center}
\vspace{1cm}

\renewcommand{\theequation}{S\arabic{equation}}
\renewcommand{\thefigure}{S\arabic{figure}}
\renewcommand{\thetable}{S\arabic{table}}
\setcounter{equation}{0}
\setcounter{section}{0}
\setcounter{figure}{0}
\setcounter{table}{0}
\section{Measurable condition for informative measurements}
\label{app0}

Here we prove Theorem 1 from the main text, i.e. 
\be
A_q\equiv i[\bar{M}_A,\bar{P}_A]=0\label{com}
\ee
for responsive measurements when $\langle a\rangle=0$ for an arbitrary weak coupling $H$, or
\be
A_c\equiv\{\bar{M}_A,\bar{P}_A\}=0\label{aco}
\ee
for informative measurements when $\langle ab\rangle=0$ for an arbitrary weak coupling $H$ 
if \begin{align}
	&\lim_{\lambda\to 0}\langle b\rangle /\lambda=i\mathrm{Tr}\bar{M}_B[\bar{P}_B\bar{P}_A,H]\nonumber\\
	&
	=i\mathrm{Tr}H\bar{P}_A[\bar{M}_B,\bar{P}_B]=\mathrm{Tr}H\bar{P}_AB_q\neq 0.
	\end{align}
for some $H$, defining $B_q$, $B_c$ analogously to $A$. In particular it means $\bar{P}_A\neq 0$, $\bar{P}_B\neq 0$, and $B_q\neq 0$.
If (\ref{com}) holds then the measurement of $A$ is responsive by definition.
On the other hand, for an arbitrary $H$,
\begin{align}
	&\lim_{\lambda\to 0}\langle a\rangle /\lambda=i\mathrm{Tr}\bar{M}_A[\bar{P}_B\bar{P}_A,H]\nonumber\\
	&
	=i\mathrm{Tr}H\bar{P}_B[\bar{M}_A,\bar{P}_A]=\mathrm{Tr}H\bar{P}_BA_q\neq 0.
	\end{align}	
Taking $H=\bar{P}_BA_q$ we have $\langle a\rangle/\lambda=\mathrm{Tr} H^2>0$, 
As $H$ Hermitian, its square is semidefinite and the trace is the sum of squares of eigenvalues,
so at leas one must be nonzero.
Consequently, $\bar{P}_B\neq 0$ and $A_q\neq 0$, 
which completes the first part of the proof by contradiction.

From the expansion
	\begin{align}
	&\lim_{\lambda\to 0}\langle ab\rangle/\lambda=i\mathrm{Tr}(\bar{M}_A\bar{M}_B)[(\bar{P}_A\bar{P}_B),H]\nonumber\\
	&=i\mathrm{Tr}H[(\bar{M}_A\bar{M}_B),(\bar{P}_A\bar{P}_B)]
	\end{align}
	but
	\begin{align}
	&2[(\bar{M}_A\bar{M}_B),(\bar{P}_A\bar{P}_B)]=\nonumber\\
	&\{\bar{M}_A,\bar{P}_A\}[\bar{M}_B,\bar{P}_B]+\{\bar{M}_B,\bar{P}_B\}[\bar{M}_A,\bar{P}_A].
	\end{align}
	It is then straightforward the if (\ref{aco}) holds for $A$ and $B$, then 
	\be
	\lim_{\lambda\to 0}
	\langle ab\rangle_\lambda/\lambda=0.\label{abh}
	\ee
	Conversely, suppose that the above is not true. By graded Jacobi identities, each Hamiltonian of the form
	$2H=FG+GF$ gives 
	\be
	\tilde{\mathcal H}=\hat{\mathcal F}\tilde{\mathcal G}
+\hat{\mathcal G}\tilde{\mathcal F}=\tilde{\mathcal F}\hat{\mathcal G}+\tilde{\mathcal G}\hat{\mathcal F},
	\ee
	and $H=i[F,G]$ gives
	\be
	\tilde{\mathcal H}=[\tilde{\mathcal G},\tilde{\mathcal F}].
	\ee
	We can take simply
	\be
	H=A_cB_q
	\ee
	and then 
	\be
	\lim_{\lambda\to 0}\langle ab\rangle/\lambda=\mathrm{Tr} A_c^2\mathrm{Tr} B_q^2
	\ee
	If follows from the fact that cross terms vanish, e.g.
	\begin{align}
	&\mathrm{Tr} A_cA_q=i\mathrm{Tr}(\bar{M}_A\bar{P}_A+\bar{P}_A\bar{M}_A)(\bar{M}_A\bar{P}_A-\bar{P}_A\bar{M}_A)\nonumber\\
	&=i\mathrm{Tr}(\bar{M}_A\bar{P}_A\bar{M}_A\bar{P}_A+\bar{P}_A\bar{M}^2_A\bar{P}_A\nonumber\\
	&-\bar{M}_A\bar{P}^2_A\bar{M}_A-\bar{P}_A\bar{M}_A\bar{P}_A\bar{M}_A),
	\end{align}
	which vanishes as the trace is cyclic and linear. As before, either $A_c=0$ or $B_q=0$,
	but the latter case is ruled out, and we get a contradiction so (\ref{aco}) must hold, and the proof is completed.$\square$

As a side note, while commutation (\ref{com}) implies joint eigenbasies, 
generally the anticommutation (\ref{aco}) implies $\mathrm{Tr}\bar{P}=0$ which gives $\sum_j\bar{p}_j=0$, known as contrast distribution.
In fact (\ref{aco}) restricts the form of $\bar{M}$ and $\bar{P}$ to a simple block structure with scaled pairs of Pauli matrices $Z$ and $X$
(see Supplemental Material \ref{appb}), and trailing 
orthogonal projections,
\be
\bar{M}=\sum_a \bar{m}_a Z_{a}+\sum_b \bar{m}_b 1_b,
\bar{P}=\sum_a \bar{p}_a X_{a}+\sum_c \bar{p}_c 1_c
\ee
where $a=(a_1,a_2)$ are different pairs of directions, $1_{b/c}$ stands for projection on $b/c$. All pairs $a$, and directions $b$, $c$ are orthogonal.

		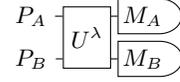
\begin{figure}
		\begin{tikzpicture}[scale=1]
			\begin{yquant*}
				qubit {} q[2];
				discard q[0,1];
				init {$P_A$} q[0];
				init {$P_B$} q[1];
				box {$U^\lambda$} (q[0],q[1]);
				dmeter {$M_A$} q[0];
				discard q[0];
				dmeter {$M_B$} q[1];
				discard q[1];
			\end{yquant*}
		\end{tikzpicture}
		
		\caption{Mutual calibration of weak measurements. For a given $H$ (or $U$) we perform the measurement for a set of preparations $P_A$, $P_B$, and measurements $M_A$, $M_B$.}
		\label{abwf}
	\end{figure}

\section{Weak measurement by a two-level detector}
\label{appa}

Here we present the proof of Theorem 2.
Suppose the detector is a qubit (space of $d=2$ dimension).
We shall use standard Pauli algebra,
\begin{equation}
\sigma_a\sigma_b=\delta_{ab}+i\epsilon_{abc}\sigma_c
\end{equation}
for $abc=1,2,3$, and
\begin{align}
&\delta_{ab}=\left\{\begin{array}{l}
1\mbox{ for }a=b,\\
0\mbox{ otherwise },\end{array}\right.,\nonumber\\
&
\epsilon_{abc}=\left\{\begin{array}{rl}
1&\mbox{ for }abc=123,231,312,\\
-1&\mbox{ for }abc=321,213,312,\\
0&\mbox{ otherwise }\end{array}\right..
\end{align}
The conventional representation of Pauli matrices is given in Supplemental Material \ref{appb}
but we do not restrict to it.
We use summation convention i.e. if a subscript appear twice we sum over it, i.e. $X_aY_a\equiv \sum_a X_aY_a$. 
We also use standard vector notation $\boldsymbol v=(v_1,v_2,v_3)$ with the symmetric dot product
\begin{equation}
\boldsymbol v\cdot\boldsymbol w=v_jw_j,
\end{equation}
length $|\boldsymbol v|=\sqrt{\boldsymbol v\cdot\boldsymbol v}$,
antisymmetric vector product $\boldsymbol u=\boldsymbol v\times\boldsymbol w$
\begin{equation}
u_a=\epsilon_{abc}v_bw_c
\end{equation}
and fully antisymmetric mixed product
\begin{equation}
[\boldsymbol u, \boldsymbol v,\boldsymbol w]=\boldsymbol u\cdot(\boldsymbol v\times\boldsymbol w)=\det(\boldsymbol u\boldsymbol v\boldsymbol w)
\end{equation}
where the last determinant is for the $3\times 3$ matrix with columns/rows $\boldsymbol u$, $\boldsymbol v$, $\boldsymbol w$.

The detector can be prepared in the state $P$,  and measured by 
an $n$-outcome measurement $M_k$, $k=1,\dots, n$ with $\sum_k M_k=1$.
The general form of the measurement is $M_{k}=\mu_{0k}+\boldsymbol \mu_k\cdot\boldsymbol \sigma$ with $\mu_{0k}\geq |\boldsymbol \mu_k|$ and $\sum_k m_{0k}=1$, $\sum_k\boldsymbol \mu_k=0$.
while initial state is $P=(1+\boldsymbol \rho\cdot\boldsymbol\sigma)/2$ with $|\boldsymbol \rho|\leq 1$ ($\rho_0=1$).
The vectors are commonly depicted in the Bloch sphere.

We have the outcome probabilities
\begin{equation}
w_k=\mathrm{Tr}M_kP=\mu_{0k}+ \boldsymbol \mu_k\cdot\boldsymbol \rho
\end{equation}
The detector will be coupled weakly, $\lambda\ll 1$, to an external system, by a generic interaction
\begin{equation}
U^\lambda=\exp\left(\lambda(\boldsymbol Q\cdot\boldsymbol\sigma+Q_0)/i\right)
\end{equation}
where $\boldsymbol Q$ is a vector of external observables and $Q_0$ is an additional observable decoupled from the detector.

The effect of this interaction on the measurement is
\begin{equation}
w^\lambda_k=\mathrm{Tr}M_k\check{\mathcal U}^\lambda(\rho P) \simeq w_k+v_{k},
\end{equation}
\begin{equation}
v_k=\boldsymbol q\cdot\boldsymbol \alpha_{k}
\end{equation}
with
\begin{equation}
\boldsymbol \alpha_{k}=\lambda\mathrm{Tr} M_k [\boldsymbol\sigma,P]/i
\end{equation}
and
\begin{equation}
\boldsymbol q=\langle \boldsymbol Q\rangle=\mathrm{Tr}\boldsymbol Q\rho.
\end{equation}
Equivalently
\begin{equation}
v_{k}=2\lambda[\boldsymbol \mu_k,\boldsymbol q,\boldsymbol \rho]
\end{equation}

If the system is subject to next measurements, there is an additional term affecting correlations after the measurement,
\begin{equation}
\mathcal K(k)=\mathrm{Tr}'M_k\check{\mathcal U}^\lambda (\cdot P)\simeq 
w_k+\boldsymbol \alpha_{k}\cdot\hat{\boldsymbol{\mathcal Q}}+\boldsymbol \alpha'_{k}\cdot\tilde{\boldsymbol{\mathcal Q}}
+\alpha'_{0k}\tilde{\mathcal Q}_0,
\end{equation}
where $\mathrm{Tr}_M$ denotes trace over the detector space,
with
\begin{align}
&\boldsymbol \alpha'_{k}=\lambda\mathrm{Tr}M_k\{\boldsymbol \sigma,P\}/2=\lambda(\mu_{0k}\boldsymbol\rho+\boldsymbol \mu_k),\nonumber\\
& \alpha'_{0k}=\lambda\mathrm{Tr}M_kP=\lambda w_k.\label{bbb}
\end{align}
Here we used the commutator-anticommutator decomposition
\be
[AB,\rho]=\{A,[B,\rho]\}/2+[A,\{B,\rho\}]/2\label{acom}
\ee
if $[A,B]=0$.

The disturbance of the system after the measurement $\rho'$ is $\sim \lambda^2$,
\begin{equation}
\mathcal K-1\simeq -\lambda^2\mathrm{Tr}' [[\cdot P,\boldsymbol Q\cdot\boldsymbol \sigma+Q_0],\boldsymbol Q\cdot\boldsymbol \sigma+Q_0]
\end{equation}
which can be reduced to
\begin{widetext}
\begin{align}
&-\lambda^2([\{\cdot, Q_a\},Q_b]\mathrm{Tr}\sigma_b [P,\sigma_a]
+[[\cdot,Q_a],Q_b]\mathrm{Tr}\sigma_b\{P,\sigma_a\})/2=
\nonumber\\
&
-\lambda^2([\{\cdot, Q_a\},Q_b]\mathrm{Tr}P[\sigma_a,\sigma_b]+[[\cdot,Q_a],Q_b]\mathrm{Tr}P\{\sigma_a,\sigma_b\})/2=
\nonumber\\
&-\lambda^2((i[\{\cdot, Q_a\},Q_b]\epsilon_{abc}+[[\cdot,Q_0],Q_c]+[[\cdot,Q_c],Q_0]])\mathrm{Tr}P\sigma_c+[[\cdot,Q_a],Q_a])
\nonumber\\
&
=
-\lambda^2((i[\{\cdot,\boldsymbol Q\times\},\boldsymbol Q]+[[\cdot,Q_0],\boldsymbol Q]+[[\cdot,\boldsymbol Q],Q_0]])\cdot\mathrm{Tr}P\boldsymbol \sigma
\nonumber\\
&
+[[\cdot,\boldsymbol Q\cdot],
\boldsymbol Q])+[[\cdot,Q_0],Q_0])=
\lambda^2((\tilde{\boldsymbol{\mathcal Q}}\tilde{\mathcal Q}_0+\tilde{\mathcal Q}_0\tilde{\boldsymbol{\mathcal Q}}+2\tilde{\boldsymbol{\mathcal Q}}\times\hat{\boldsymbol{\mathcal Q}})
\cdot\boldsymbol \rho+\tilde{\boldsymbol{\mathcal Q}}\cdot\tilde{\boldsymbol{\mathcal Q}}+\tilde{\mathcal Q}^2_0)
\end{align}
\end{widetext}
where we used $\sigma_0=1$ in the summation in the first two lines.
Our aim is to arrange a combination of at least 3 initial states $P_{1,2,3}$ and 
 and reassignment of  the outcomes of  a single 4-outcome measurements, $M_k$, that  $\bar{M}=\sum_k\bar{m}_kM_k$ and $\bar{P}=\sum_j\bar{p}_j$
 satisfy $\{\bar{M},\bar{P}\}=0$,
 and
to recover standard weak measurement vector
\begin{equation}
\bar{\boldsymbol \alpha}=\mathrm{Tr}\bar{M}[\boldsymbol \sigma,\bar{P}]/i\label{aaa}
\end{equation}
and $A=\hat{\boldsymbol \alpha}\cdot\boldsymbol{Q}$.
It turns out that the sole knowledge of the $w_k$ does not suffice to determine $\mu_{0k}$. 
We will need help from a weak measurement, the corrections  $v_k$ used here for calibration, and still we do not need any knowledge about the external state.

\begin{figure}
\begin{center}
\includegraphics[scale=1]{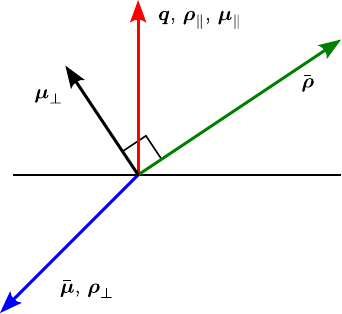}
\end{center}
\caption{The relative placement of vectors in the calibration of weak measurement
$\boldsymbol q$ is parallel to $\boldsymbol \rho_\parallel$ and  $\boldsymbol \mu_\parallel$.
Vector $\boldsymbol \mu_\perp$ is perpendicular to the plane of differences of $\boldsymbol \rho_j$ vectors.
Vector $\bar{\boldsymbol \rho}$ lies in this plane, at the same time in the plane spanned by $\boldsymbol \mu_\perp$  and $\boldsymbol q$.
Vectors $\boldsymbol \rho_\perp$  and $\bar{\boldsymbol \mu}$ are parallel and perpendicular to $\boldsymbol q$ and $\boldsymbol \mu_\perp$ 
and $\boldsymbol \rho_\perp$.}
\label{pvm}
\end{figure}

We have the probability matrices
$w_{kj}$, $v_{kj}$
obtained for the same coupling to an external system and its observables $\boldsymbol Q$, $Q_0$.
The last outcome $k=4$ is not independent in the matrices $w$ and $w$ from normalization of $M$,
\begin{equation}
\sum_{k=4}^3w_{kj}=1,\;\sum_{k=1}^4v_{kj}=0
\end{equation}
Let us then take 3 (first) rows,
or multiply $w$ and $v$ by some  $3\times 4$ reduction matrix.
We have rk $v=2$ as generic measurements/states corresponding to Bloch vectors parallel to $\boldsymbol q$ belong to $\mathrm{ker}\;v$.
This follows from our critical assumption that the dimension of the detector is $d=2$. If $\det v\neq 0$, and so rk $v=3$ then
 the detection process involves also other degrees of freedom or a non-unital quantum channel.
We define
\begin{equation}
\boldsymbol m_\parallel \in \mathrm{ker}\;v^T,\:\boldsymbol p_\parallel \in \mathrm{ker}\;v
\end{equation}
These vectors can be defined in a generic way, taking some constant vectors (to be determined e.g. by optimization)
$
\boldsymbol a$, $\boldsymbol b$, $\boldsymbol c$, $\boldsymbol d$ so that
\begin{equation}
m_{\parallel k}=\epsilon_{kjn}v_{jr}a_rv_{ns}b_s,\:
p_{\parallel j}=\epsilon_{jkn}c_rv_{rk}d_sv_{sn},
\end{equation}
It allows to construct $M_\parallel=m_{\parallel k}M_k$ and $P_\parallel=p_{\parallel j}P_j$ with 
\begin{equation}
\boldsymbol \mu_\parallel=m_{\parallel k}\boldsymbol \mu_k,\:\boldsymbol \rho_\parallel=p_{\parallel j}\boldsymbol \rho_j,
\end{equation}
which are both parallel to $\boldsymbol q$, and $\rho_{\parallel 0}=\sum_j p_{\parallel j}$. These and next vectors are then unique up to scaling.
Now
\begin{equation}
\boldsymbol m_\perp=\boldsymbol w_1\times \boldsymbol w_2+\boldsymbol w_2\times \boldsymbol w_3+\boldsymbol w_3\times \boldsymbol w_1
\end{equation}
where vectors $\boldsymbol w_j$ are columns of $w=(\boldsymbol w_1\boldsymbol w_2\boldsymbol w_3)$. It defines vector
$
\boldsymbol \mu_\perp=m_{\perp k}\boldsymbol \mu_k$ perpendicular to the plane spanned by $\boldsymbol \rho_1-\boldsymbol \rho_3$ and $\boldsymbol \rho_2-\boldsymbol \rho_3$.
Then our final initial state coefficients are defined
\begin{equation}
\bar{p}_j=\sum_n\epsilon_{jln}m_{\perp k}v_{kl}
\end{equation}
giving $\bar{P}=\bar{p}_jP_j$ and
$
\hat{\boldsymbol{\rho}}=\bar{p}_j\boldsymbol \rho_j.
$
The vector $\bar{\boldsymbol{\rho}}$ lies in the plane spanned by $\boldsymbol q$  and $\boldsymbol \mu_\perp$ while perpendicular to $\boldsymbol \mu_\perp$.
We need auxiliary $p_{\perp j}=\sum_n\epsilon_{jln}m_{\parallel k}w_{kl}$
giving 
\begin{equation}
\boldsymbol \rho_\perp=p_{\perp j}\boldsymbol \rho_j
\end{equation}
which is perpendicular to $\boldsymbol \mu_\parallel$, $\boldsymbol \mu_\perp$, and $\bar{\boldsymbol \rho}$.
The final measurement coefficients are given by
\begin{equation}
m'_k=\epsilon_{kjn}v_{jr}p_{\perp r}w_{ns}\bar{p}_s
\end{equation}
with 
\begin{equation}
\bar{\boldsymbol \mu}=\boldsymbol \mu'=m'_k\boldsymbol \mu_k 
\end{equation}
and $\mu'_0=m'_k \mu_{0k}$. It can be nonzero but fortunately we can determine as
\begin{equation}
m'_kw_{kj}p_{\parallel _j}=\mu'_0\rho_{\parallel 0}+\bar{\boldsymbol \mu}\cdot\boldsymbol \rho_\parallel=\mu'_0\rho_{\parallel 0}
\end{equation}
as $\bar{\boldsymbol \mu}$ lies in the plane spanned by $\boldsymbol q$ and $\boldsymbol \rho_\perp$ and perpendicular to $\bar{\boldsymbol \rho}$ 
and so it is parallel to $\boldsymbol \rho_\perp$ and  perpendicular to $\boldsymbol q$ and $\boldsymbol \rho_\parallel$.
The final construction of the measurement reads then
\begin{equation}
\bar{M}=m'_kM_k-\mu'_0(M_1+M_2+M_3+M_4)
\end{equation}
for 
\be
\mu'_0=m'_kw_{kj}p_{\parallel j}/\sum_j p_{\parallel j}.
\ee
 since $\sum_k M_k=1$, but we need to check if the denominator is nonzero. Equivalently
\begin{equation}
\bar{m}_k=\left\{\begin{array}{l}
m_k'-\mu'_0\mbox{ for }k=1,2,3\\
-\mu'_0\mbox{ for }k=4
\end{array}\right.
\end{equation}
The configuration of vectors is depicted in Fig. \ref{pvm}, showing also their uniqueness.

The matrices $w$ and $v$ allow to find the traceless perpendicular operators $\bar{M}$ and $\bar{P}$ but not the magnitude of the Bloch vectors,
$|\bar{\boldsymbol \rho}|$ and $|\bar{\boldsymbol \mu}|$. We can only make an estimate, i.e.
\begin{equation}
\mathrm{max}_j |\bar{m}_kw_{kj}|=\mathrm{max}_j|\bar{\boldsymbol \mu}\cdot\boldsymbol \rho_j|\leq |\bar{\boldsymbol \mu}|=
\left|\bar{m}_k\boldsymbol \mu_k\right|\leq \sum_k|\bar{m}_k|
\end{equation}
since $|\boldsymbol \mu_k|,|\boldsymbol \rho_j|\leq 1$. Similarly 
\begin{equation}
\mathrm{max}_k |w_{kj}\bar{p}_j|=\mathrm{max}_k|\bar{\boldsymbol \rho}\cdot\boldsymbol \mu_k|\leq |\bar{\boldsymbol \rho}|=
\left|\bar{p}_j\boldsymbol \rho_j\right|\leq \sum_j|\bar{p}_j|
\end{equation}
In general, the inequalities leave only a certain interval. We can narrow it taking more measurements and preparations which are accurate
(projections). The genericity in the sens of Haar measure follows from the arbitrariness of directions $\boldsymbol q$, $\boldsymbol \rho_j$ and $\boldsymbol \mu_k$
which are distributed uniformly so that degenerate cases are accidental. $\square$

A natural example is a measurement along vectors
\begin{align}
&\boldsymbol \mu_1=(1,-1,-1)/4\sqrt{3},\nonumber\\
&\boldsymbol \mu_2=(-1,1,-1)/4\sqrt{3},\nonumber\\
&\boldsymbol \mu_3=(-1,-1,1)/4\sqrt{3},\nonumber\\
&\boldsymbol \mu_4=(1,1,1)/4\sqrt{3}
\end{align}
and $\mu_{0k}=1/4$.
The operators are 4 scaled projections pointing at the vertices of regular tetraherdon as sum up to identity. Such a measurement can be easily realized with an auxiliary qubit by an appropriate entanglement and rotations.

In this case the matrices read
\begin{equation}
w=\frac{1}{4}\begin{pmatrix}
1&1&1\\
1&1&1\\
1&1&1\\
1&1&1
\end{pmatrix}+\frac{1}{4\sqrt{3}}\left(
\begin{array}{rrr}
1&-1&-1\\
-1&1&-1\\
-1&-1&1\\
1&1&1
\end{array}\right)
\end{equation}
and
\begin{equation}
v=\frac{\lambda q}{2\sqrt{3}}
\left(
\begin{array}{rrr}
-1&-1&0\\
1&1&0\\
-1&1&0\\
1&-1&0\end{array}\right)
\end{equation}
Excluding the fourth row, we can take 
$\boldsymbol a=\boldsymbol c=\boldsymbol e_1$, $\boldsymbol b=\boldsymbol e_2$, $\boldsymbol d=\boldsymbol e_3$ we obtain
\begin{equation}
\boldsymbol m_{\parallel}=(1,1,0,0)(\lambda q)^2/6,\:\boldsymbol p_{\parallel}=(0,0,-1)(\lambda q)^2/6
\end{equation}
with $\rho_{\parallel 0}=-(\lambda q)^2/24$
and 
\begin{equation}
\boldsymbol m_{\perp}=(1,1,1,0)/12,\:\boldsymbol \mu_{\perp}=-(1,1,1)/36\sqrt{3}
\end{equation}
$\mu_{\perp 0}=1/16$,
while
\begin{align}
&\bar{\boldsymbol p}=(1,1,-2)\lambda q/2^3 3\sqrt{3}=\bar{\boldsymbol \rho},\nonumber\\
&\boldsymbol p_\perp=(1,-1,0)(\lambda q)^3/2^2 3\sqrt{3}=\boldsymbol \rho_\perp
\end{align}
and finally 
\begin{align}
&\bar{\boldsymbol m}=(1,-1,0,0)(\lambda q)^5/2^{6} 3^4,\nonumber\\
&\bar{\boldsymbol \mu}=(1,-1,0)(\lambda q)^5/2^{7} 3^4\sqrt{3}
\end{align}
We have the inequalities for  $|\bar{\boldsymbol \rho}|=\sqrt{6}\lambda q/2^3 3\sqrt{3}$ and $|\bar{\boldsymbol \mu}|$, giving
\begin{align}
&1/\sqrt{3}\leq 2^3 3\sqrt{3}|\bar{\boldsymbol \rho}|/\sqrt{6}|\lambda q|\leq 4,
\nonumber\\
&
1/2\sqrt{3}\leq 2^{6} 3^4|\bar{\boldsymbol \mu}|/|\lambda q|^5\leq 2
\end{align}
Both $\bar{P}$ and $\bar{M}$ can be arbitrarily rescaled as a single weak measurement get information about the measurement system multiplied
by an unknown factor. The scaling does not affect the disturbance, which does not depend on the measurements and can be simply estimated 
by the relative difference in the results of next measurements with and without the weak measurements.

Instead of a single $4-$outcome measurement one can take generic two measurements with 2 and 3 outcomes, or three measurements with 2 outcomes.

\section{Classical weak measurements}
\label{apcl}
Classical measurements essentially replace:
\begin{itemize}
\item Hilbert space by phase space,
\item operators by functions in phase space,
\item maps by linear functionals,
\item commutator by Poisson bracket,
\item anticommutator by multiplication,
\item trace by phase space integral.
\end{itemize}
Let us formalize it. We have a phase space consisting of conjugate variables,
here for simplicity position $x$ and momentum $q$ (we can add more pairs).
The functions in phase space have the form $A(x,q)$ or states $\rho(x,q)$. 
The averages are obtained by integration over the phase space
$\langle A\rangle=\int A\rho \d x\d q$, in particular $\langle 1\rangle=1$ from normalization.
For the system-meter interaction we need to add the meter space
$\bar{x}$, $\bar{q}$. We recall Poisson bracket
\be
(A,B)=\sum_j\left(\frac{\partial A}{\partial x_j}\frac{\partial B}{\partial q_j}-\frac{\partial A}{\partial q_j}\frac{\partial B}{\partial x_j}\right).
\ee
For a simple system-meter space $x_1=x$, $x_2=\bar{x}$, $q_1=q$, $q_2=\bar{q}$.
The useful identity are $(A,B)=-(B,A)$, $(AB,C)=A(B,C)+(A,C)B$, and $((A,B),C)+((B,C),A)+((C,A),B)=0$ (Jacobi).
We define the maps corresponding to the quantum ones
\be
\hat{\mathcal A}\rho=A\rho,\:\tilde{A}\rho=(A,\rho).
\ee
The map $\hat{\mathcal A}$ replaces the anticommutator by a multiplication by a function, while $\tilde{\mathcal A}$ 
replaces the commutator by a Poisson bracket, using the dimensional convention $\hbar=1$. 
We can also write 
\begin{align}
&\tilde{\mathcal C}=\hat{\mathcal A}\tilde{\mathcal B}+\hat{\mathcal B}\tilde{\mathcal A}=
\tilde{\mathcal A}\hat{\mathcal B}+\tilde{\mathcal B}\hat{\mathcal A},\nonumber\\
&\tilde{\mathcal D}=[\tilde{\mathcal B},\tilde{\mathcal A}]
\end{align}
if 2$C=AB+BA$, and $D=(B,A)$, and $\hat{\mathcal A}\tilde{\mathcal A}=\tilde{\mathcal A}\bar{\mathcal A}$, analogous to the quantum ones.

Now the coupling Hamiltonian is $H(x,q,\bar{x},\bar{q})$, and the unitary map is $\check{\mathcal U}^\lambda=\exp\lambda\mathcal H$.
The meter's intial preparations is $P(\bar{x},\bar{q})\geq 0$ and measurement $M(a;\bar{x},\bar{q})\geq 0$ with $\int P\d\bar{x}\d\bar{q}=1$ and $\int M(a)\d a=1$,
which agrees with classical probability. The tensor product is a simple product $P\rho$ within different phase spaces.
Our quantum instrument is replaced by the classical one
\be
\mathcal K(a)=\int  M(a)\check{\mathcal U}^\lambda(\rho P) \d \bar{x}\d \bar{q},
\ee
and the final measurement can be defined within the system's space $M(a;x,q)$ and $\mathcal K(a)\rho=\int M(a)\rho \d x\d q$.
It is clear that classical and quantum instruments are analogues, except disturbance. 
In the classical case
it is $\mathcal K\rho=\int \check{\mathcal U}^\lambda(\rho P)\d\bar{x}\d\bar{q}$ but now one can in principle construct a completely noninvasive measurement
with $\lambda=1$, namely, 
\be
P=\delta(\bar{x})\delta(\bar{q}),\; H=\bar{q} A,\; M(a)=\delta(a-\bar{x}).
\ee
The evolution acts as 
\be
\tilde{\mathcal H}(\rho P)=\bar{q}(A,\rho)P-A\partial P/\partial \bar{x}.
\ee
The first term vanishes due to $\delta(\bar{q})$ in $P$ while the other has $\delta(\bar{q})$ again. Therefore
\be
\check{\mathcal U}(\rho P)=\delta(\bar{x}-A)\delta(\bar{q})\rho,
\ee
and $\mathcal K=1$ but $\mathcal K(a)=\delta(a-A)$ so it retrieves the value of $A$ without any disturbance.
We cannot copy it to the quantum case due to uncertainty principle between $\bar{x}$ and $\bar{q}$.

Let us now find the classical counterpart of weak measurement. Similarly as in the quantum case, we define $\bar{M}=\int aM(a)\d a$
and $\bar{P}=\sum_j\bar{p}_j P_j$, to get
\be
\bar{\mathcal K}=\int \bar{M} \check{\mathcal U}^\lambda(\cdot \bar{P})\d\bar{x}\d\bar{q}
\ee
which defines averages
\be
\langle \cdots \mathcal K\cdots\rangle=\sum_j\bar{p}_j\langle \cdots a\cdots \rangle\equiv \langle \cdots a \cdots\rangle.
\ee
We shift $a$ to get $\langle a\rangle=0$ for $\lambda =0$.
Expanding $\check{\mathcal U}^\lambda=1+\lambda\tilde{\mathcal H}$ for $\lambda\ll 1$, we get
\be
\bar{\mathcal K}\simeq \lambda(\hat{\mathcal A}+\tilde{\mathcal A}')
\ee
with the informative and responsive part, respetively,
\be
 A=\int (\bar{P},\bar{M})H\d\bar{x}\d\bar{q},\:A'=\int \bar{M}\bar{P}H\d\bar{x}\d\bar{q},
\ee
because
\begin{align}
&\tilde{\mathcal H}(\rho \bar{P})=(H,\rho \bar{P})=\nonumber\\
&(H,\rho)\bar{P}+\rho (H,\bar{P})=(H,\rho)_1 \bar{P}+\rho (H,\bar{P})_2,
\end{align}
where subscripts $1,2$ indicate the bracket restricted to the system and meter, respectively.
Then intergation with $\bar{M}$ commutes with the first bracket, while the second can be moved by parts to $\bar{M}$,
\be
\int \bar{M}(H,P)_2\d\bar{x}\d\bar{q}=\int H(\bar{P},\bar{M})\d\bar{x}\d\bar{q}.
\ee
As in the quantum case $\langle a\rangle\simeq \lambda\langle A\rangle$ because the Poisson bracket is integrated out to zero,
and we shall call the case $A\neq 0$, $A'=0$, informative measurement and $A=0$, $A'\neq 0$, responsive measurement.
The Theorem 1 holds for classical measurements too, giving the measureble criterion of $A'=0$ whenever $\bar{M}\bar{P}=0$.
Nevertheless it is possible to realize a responsive classical measurement by
$\bar{P}=e^{-\bar{x}^2-\bar{q}^2}/\pi$, $H=2\bar{x}A'$, $\bar{M}=\bar{x}$. Note that the meter needs some initial variance of $\bar{x}$ to
keep finite $H$. There is a finite disturbance $\mathcal K=\exp\lambda^2\tilde{\mathcal A}^{\prime 2}$, 
because $\tilde{\mathcal H}=2\bar{x}\tilde{\mathcal A}'+2\hat{\mathcal A}'\partial/\partial \bar{q}$
and the derivative over $\bar{q}$ is integrated out. It cannot be reduced due to the normalization of the internal meter's noise,
\be
\langle a^2\rangle\geq\frac{\langle \bar{x}\bar{M}\rangle^2}{\langle \bar{x}^2\rangle}=1/2.
\ee

We shall show that classical responsive measurements violate (\ref{lgg2}) if substituting $a,b\to \tilde{\mathcal A}'$ and $c\to C$. Let us write (\ref{lgg2}) in a slightly different form
\be
\langle a+b|c\rangle^2/4\leq \langle ab|c\rangle.\label{aabb}
\ee
The condition will be the point measurement at $x=z$, $C=\delta(x-z)$, the system state $e^{\alpha(x)}=\int \rho(x,q)\d q$ , and $A'=q$.
Strictly speaking, the condition should cover an infinitesimal interval $\Delta z>0$ around $z$ 
but it would enter simply as a common prefactor canceling in conditional averages in the limit $\Delta z\to 0$.
Then, with $\alpha\equiv \alpha(z)$,
$\langle C\rangle=e^{\alpha}$,
$
\langle C\tilde{\mathcal A}'\rangle=-e^{\alpha}\alpha'
$
and
$
\langle C\tilde{\mathcal A}^{\prime 2}\rangle=e^{\alpha}(\alpha^{\prime 2}+\alpha''$ so putting it together, the inequality reads
$\alpha^{\prime 2}\leq \alpha^{\prime 2}+\alpha''$.

which is false, whenever $\alpha''<0$, meaning that $x$ marginal of $\rho$ is log-concave. For instance, $\rho=e^{-x^2-q^2}/\pi$ has $\alpha''=-2$. 

The origin of the failure is the attempt to interpret the meter-detector interaction as an information gain.
However, the result is not an information  from the system, which is encoded in $\bar{q}$, but
the response of the system correlated with the meter's outcome. The finite overall distrubance in not a fundamental issue because, taking $\bar{M}=\bar{q}$
instead,
we have the desired informative measurement.

\section{Violation with position and momentum}
\label{apw}

We can find violation of (\ref{lgg2}) in the form (\ref{aabb}) with position and momentum operators, $X$, $Q$.
We have $[X,Q]=i$ (convention $\hbar=1$) with eigenstates $X|x\rangle=x|x\rangle$ and $Q|q\rangle=q|q\rangle$, normalized $\langle x|x'\rangle=\delta(x-x')$, $\langle q|q'\rangle=\delta(q-q')$, and 
\be
\int e^{iqx}|x\rangle\frac{\d x}{\sqrt{2\pi}}=|q\rangle,\;\int e^{-iqx}|q\rangle\frac{\d q}{\sqrt{2\pi}}=|x\rangle.
\ee
In this basis $Q|x\rangle=i\partial_x|x\rangle$ and $X|q\rangle=-i\partial_q|q\rangle$.

Take now a state 
\be
|\psi\rangle=\int \psi(x)|x\rangle \d x,
\ee
so $\rho=|\psi\rangle\langle \psi|$
and make informative measurements $A=B=Q$ and condition by $C=|z\rangle\langle z|$, projection at $x=z$,
again ignoring infinitesimal interval $\Delta z$.
Then $\langle C\rangle=|\psi|^2$ with the shorthand $\psi\equiv \psi(z)$,
\be
\hat{\mathcal Q}\rho=\int i(\psi(x)\psi^{\ast\prime}(y)-\psi'(x)\psi^\ast(y))|x\rangle\langle y|\d x \d y/2,
\ee
which gives
$\langle C\hat{\mathcal Q}\rangle=\mathrm{Im}\psi'\psi^\ast$,
and finally
\begin{align}
&\hat{\mathcal Q}^2\rho=\int (2\psi'(x)\psi^{\ast\prime}(y)-\psi(x)\psi^{\ast\prime\prime}(y)-\psi''(x)\psi^\ast(y))\times\nonumber\\
&|x\rangle\langle y|\d x \d y/4,
\end{align}
giving
\be
\langle C\hat{\mathcal Q}^2\rangle=(|\psi'|^2-\mathrm{Re}\psi''\psi^\ast)/2.
\ee
Assuming the polar form $\psi=e^{\alpha+i\beta}$ with real $\alpha$ and $\beta$ we have
\begin{align}
&\langle C\rangle=e^{2\alpha},\nonumber\\
&\langle C\hat{\mathcal Q}\rangle=\beta'e^{2\alpha},\nonumber\\
&\langle C\hat{\mathcal Q}^2\rangle=(\beta^{\prime 2}-\alpha''/2)e^{2\alpha}.
\end{align}
We get the inequality (\ref{aabb}) in the form
\be
\beta^{\prime 2}\leq \beta^{\prime 2}-\alpha''/2.
\ee
It is  violated if 
\be
\alpha''> 0,
\ee
meaning that the amplitude is log-convex.

Let us take a superposition a two Gaussians, a cat state
\be
\psi(x)=\frac{e^{-(x-a)^2/2}+e^{-(x+a)^2/2}}{\sqrt{2\sqrt{\pi}(1+e^{-a^2})}}.
\ee
Then $
\alpha'=a\tanh(ax)-x$
and
$
\alpha''=a^2-1
$
at $x=z=0$, which is positive if $a^2>1$.

We can confirm this analyzing Wigner function.
We recall standard definition of dimensionless Wigner function
\begin{align}
&W(x,q)=\int\langle x-y|\rho|x+y\rangle e^{2iqy}\frac{\d y}{\pi}=\nonumber\\
&\int \frac{\d x' \d q'}{(2\pi)^2}e^{iqx'-ixq'}\langle e^{iXq' - iQx'}\rangle.
\end{align}
Our maps translate to operations on Wigner function,
\begin{align}
&\hat{\mathcal X}\rho \to xW,\;
\hat{\mathcal Q}\rho \to qW,\nonumber\\
&\tilde{\mathcal X}\rho \to \partial_q W,\;
\tilde{\mathcal Q}\rho \to -\partial_x W,\nonumber\\
&\langle x|\rho|x\rangle=\int W(x,q)\d q,\;
\langle q|\rho|q\rangle=\int W(x,q)\d x,\nonumber\\
&\mathrm{Tr}\rho\to \int W\d x \d q.
\end{align}
For the cat state we have
\be
\pi W(x,q)=\frac{e^{-x^2-q^2}}{1+e^{-a^2}}(e^{-a^2}\cosh(2ax)+\cos(2aq)).
\ee
Restricting to $x=z=0$, we get
\be
\pi W(0,q)=\frac{e^{-q^2}}{1+e^{-a^2}}(e^{-a^2}+\cos(2aq)).
\ee
so
\begin{align}
&\int W(0,q)\sqrt{\pi}\d q=\frac{2}{e^{a^2}+1},\nonumber\\
&\int W(0,q)q^2\sqrt{\pi}\d q=\frac{1-a^2}{e^{a^2}+1},
\end{align}
and finally
\be
\frac{\langle \delta(x) q^2\rangle_W}{\langle\delta(x)\rangle_W}=-\alpha''/2,
\ee
with $W$ acting as a quasiprobability,
in agreement with the direct calculation of $\alpha''$.
For the Gaussian state, $a=0$, responsive measurements violate (\ref{aabb})  due to classical-quantum equivalence.

Interestingly, Fock states do not violate (\ref{aabb}).
They have the wavefunctions
\be
\psi_n(x)=N_ne^{-x^2/2}H_n(x),
\ee
with Hermite polynomials $H_n$ and normalization $N_n$.
We shall show that they are log-concave, finding
\be
\alpha=\ln H_n-x^2/2+\ln N_n,
\ee
so
$
\alpha'=H_n'/H_n-x,
$
and
\be
\alpha''=H_n''/H_n-H_n^{\prime 2}/H_n^2-1.
\ee
However,
\be
H_n^{\prime 2}+H_n^2\geq H_n'' H_n,
\ee
since
$H''_n=2n H'_{n-1}=4n(n-1)H_{n-1}$ from the property $H'_n=2nH_{n-1}$, and due to Turan inequality
$H_{n-1}^2\geq H_n H_{n-2}$.

	\section{IBM Quantum circuits}
	\label{appb}
	The Pauli matrices read in $|0\rangle$, $|1\rangle$ basis
	\begin{align}
	&X=\sigma_1=\begin{pmatrix}
		0&1\\
		1&0\end{pmatrix},\:Y=\sigma_2=\begin{pmatrix}
		0&-i\\
		i&0\end{pmatrix},\nonumber\\
	&Z=\sigma_3=\begin{pmatrix}
		1&0\\
		0&-1\end{pmatrix},\:
	I=\sigma_0=\begin{pmatrix}
		1&0\\
		0&1\end{pmatrix}.\label{pauli}
	\end{align}
	We use expansion
	$V_\theta=\exp \theta V/2i=\cos\theta/2-iV\sin\theta/2$ if $V^2=1$ and 
	\be
	Z_\theta=\begin{pmatrix}
	e^{-i\theta/2}&0\\
0&e^{i\theta/2}\end{pmatrix}
\ee
	
	The basic single-qubit gate is
	\be
	X_+=X_{\pi/2}=(I-iX)/\sqrt{2}=\begin{pmatrix}
		1&-i\\
		-i&1\end{pmatrix}/\sqrt{2},
	\ee
	
	The observable  $Z$, can be changed by rotation on the qubit,
	\be
	X=Y_+ZY_-,\:Y=X_-ZX_+
	\ee
	expressed by native gates  $Y_\pm=Z_\pm X_+ Z_\mp$ and $X_-=ZX_+Z$. 
	The two-qubit $(ZZ)_\theta\equiv RZZ(\theta)$ gate reads
	\begin{equation}
		\begin{pmatrix}
			e^{i\theta/2}&0&0&0\\
			0&e^{-i\theta/2}&0&0\\
			0&0&e^{-i\theta/2}&0\\
			0&0&0&e^{i\theta/2}
		\end{pmatrix}
	\end{equation}
	in the basis $|00\rangle$, $|01\rangle$, $|10\rangle$, $|11\rangle$.

\section{Imperfect detection scheme}
\label{appc}

	We shall discuss deviation from the ideal instrument for imperfect preparations, measurements and single-qubit gates.
	Let the initial state
	$P=(1-\epsilon)|0\rangle\langle 0|+\epsilon |1\rangle\langle 1|$, overrotation of $X_+$ in $Y_\pm$ by an angle $\alpha$, 
	$X_+\to X_{\pi/2+\alpha}$ and the last $X_+$ by $\beta$, and
	measurement $M_0\equiv M_+=\eta +(1-\omega)|0\rangle\langle 0|$.
	The initial state is now
	\begin{align}
	&2P_\pm=Z_\pm X_{\pi/2+\alpha}(1+(1-2\epsilon)Z)X^\dag_{\pi/2+\alpha}Z^\dag_\pm=\nonumber\\
		&1-(1-2\epsilon)Z_\pm X_\alpha YX^\dag_\alpha Z^\dag_\pm=\nonumber\\
		&1-(1-2\epsilon)Z_\pm(\cos\alpha-i\sin\alpha X)YZ^\dag_\pm=\nonumber\\
		&1-(1-2\epsilon)Z_\pm(Y\cos\alpha+Z\sin\alpha)Z^\dag_\pm=\nonumber\\
		&1-(1-2\epsilon)(Z\sin\alpha\mp X\cos\alpha).	
	\end{align}
	Then
	\begin{align}
	&2(ZZ)_\theta (P_\pm\rho) (ZZ)^\dag_\theta=\nonumber\\
	&P_\pm\rho(\cos \theta+1)-i[ZZ,P_\pm\rho]\sin\theta\nonumber\\
	&+ZZ(P_\pm\rho)ZZ(1-\cos \theta)
	\end{align}
	so the disturbance reads
	\begin{align}
	&2\rho'=2\mathrm{Tr}_M(ZZ)_\theta (P_\pm\rho) (ZZ)^\dag_\theta=\nonumber\\
	&\rho(1+\cos \theta)+i[Z,\rho](1-2\epsilon)\sin\alpha\sin\theta\nonumber\\
	&+Z\rho Z(1-\cos\theta)
	\end{align}
	while the difference of measurements
	\begin{align}
	&\bar{M}=M_+-M_-=2M_+-1=\nonumber\\
	&2\eta-1+2(1-\omega)X^\dag_{\pi/2+\beta}|0\rangle\langle 0|X_{\pi/2+\beta}\nonumber\\
	&=2\eta-\omega+(1-\omega)X^\dag_{\beta}YX_\beta=\nonumber\\
	&=2\eta-\omega+(1-\omega)Y(\cos\beta-iX\sin\beta)=\nonumber\\
	&=2\eta-\omega+(1-\omega)(Y\cos\beta-Z\sin\beta)
	\end{align}
	as $2|0\rangle\langle 0|=1+Z$.
	Defining 
	\begin{align}
	&\bar{\mathcal K}_\pm\rho=(\mathcal K_\pm(+)-\mathcal K_\pm(-))\rho\nonumber\\
	&=\mathrm{Tr}' \bar{M}(ZZ)_\theta (P_\pm\rho)(ZZ)^\dag_\theta
	\end{align}
	we get, using (\ref{acom}),
	\begin{align}
	&\bar{\mathcal K}_\pm\rho=\nonumber\\
	&(2\eta-\omega+(1-2\epsilon)(1-\omega)\sin\alpha\sin\beta)(1+\cos \theta)\rho/2\nonumber\\
	&+i[Z,\rho]\sin\theta((2\eta-\omega)(1-2\epsilon)\sin\alpha+(1-\omega)\sin\beta)/2\nonumber\\
	&\pm\sin\theta\{Z,\rho\}\cos\alpha\cos\beta(1-2\epsilon)(1-\omega)/2+\nonumber\\
	&(2\eta-\omega+(1-2\epsilon)(1-\omega)\sin\alpha\sin\beta) Z\rho Z(1-\cos\theta)/2
	\end{align}
    which leads to only to rescaling
	\begin{equation}
	2\bar{\mathcal K}=\bar{\mathcal K}_{+}-\bar{\mathcal K}_{-}=2\hat{\mathcal Z}(1-2\epsilon)(1-\omega)\cos\alpha\cos\beta\sin\theta.
	\end{equation}
	The disturbance can be expressed
	\be
	\mathcal K=1-f\hat{\mathcal Z}\rho+g\tilde{\mathcal Z}^2\label{inv}
	\ee
	with  $f=(1-2\epsilon)\sin\alpha\sin\theta$ and
	$g=(1-\cos\theta)/4$.
	We can also write
	\begin{align}
	&\bar{\mathcal K}_\pm=(2\eta-\omega+(1-2\epsilon)(1-\omega)\sin\alpha\sin\beta)\nonumber\\
	&-\tilde{\mathcal Z}\sin\theta((2\eta-\omega)(1-2\epsilon)\sin\alpha+(1-\omega)\sin\beta)/2\nonumber\\
	&\pm\hat{\mathcal Z}\sin\theta\cos\alpha\cos\beta(1-2\epsilon)(1-\omega)+\nonumber\\
	&\tilde{\mathcal Z}^2(2\eta-\omega+(1-2\epsilon)(1-\omega)\sin\alpha\sin\beta)(1-\cos\theta)/8
	\end{align}
	
	In the case of noisy dichotomic measurement the invasiveness reads
	\be
	\mathcal A=1-f\tilde{A}+g\tilde{\mathcal A}^2
	\ee
	as given by (\ref{inv}).
	
	Let all observables $A$, $B$, $C$, be dichotomic.
	Note that the strength of the last measurements is irrelevant.
	For two weak measurements, first $A\to B$ followed by $C$, we have four possibilities
	$\langle C\hat{\mathcal B}\mathcal A\rangle$, 
	$\langle C\mathcal B\hat{\mathcal A}\rangle$,
	$\langle C\mathcal B\mathcal A\rangle$, $\langle \hat{\mathcal B}\mathcal A\rangle$.
	Here $\mathcal O$ means that $O$ is measured with ignored outcomes, so it may affect other observables
	as in definition of $\mathcal K$.
	
	To find corrections of finite invasiveness to the measurement $A\to B\to C$ with $A=B=Z$, we note that
	$2C=1-X\sin \psi-Z\cos \psi$, $2\rho=1+X\sin \psi-Z\cos \psi$, $2\hat{\mathcal Z}C=Z-\cos \psi$,
	$\tilde{\mathcal Z}^2C=-2X\sin \psi$, $\tilde{\mathcal Z}^4 C=-8X\sin \psi $ so
	 the only change is for the last measurement alone $\langle C\rangle$ which gets corrections from the preceding weak ones.
	Since $\tilde{\mathcal Z}C=2iZX\sin \psi=-2Y\sin \psi$, for noisy measurements we have a single expression
	\be
	\langle C\mathcal B\mathcal A\rangle=\cos^2 \psi+2(g_A+g_B-4g_Ag_B+f_Af_B)\sin^2 \psi.
	\ee
	Note that the projective measurement means $g=1/4$ and $f=0$ which results in the loss of coherence along $Z$ and 
	the result $1-(\sin^2 \psi)/2$.
	
\section{Additional data}
\label{appd}

 	\begin{table}
		\begin{tabular}{*{8}{c}}
			\toprule
			$C$&$A$&$B$&$eCA$&$eCB$&$eC$&$eA$&$eB$\\
129&118&128&0.17&0.14&0.77&0.6&1\\
83&82&96&0.17&0.11&4.7&1.2&0.89\\
141&140&142&0.088&0.12&0.46&0.61&0.74\\
104&103&105&0.13&0.14&0.76&1&0.59\\
125&117&126&0.14&0.15&2.3&0.56&0.57\\
49&38&50&0.12&0.17&1&0.72&0.37\\
14&13&15&0.16&0.077&1.9&0.66&0.82\\
63&62&64&0.2&0.23&1&2.2&0.63\\
70&69&71&0.24&0.19&2.1&1&0.72\\
67&57&68&0.15&0.16&0.93&0.59&0.48\\
			\bottomrule
		\end{tabular}
		\caption{Calibration data for the tested qubits from \texttt{ibm\_kingston}, main (system) $C$,
			auxiliary (meters) $A$ and $B$, errors of $RZZ$ gates between qubit $C$ and
			$A/B$, and readout errors on qubits $C$, $A$, $B$. Error unit $10^{-2}$.}
		\label{fffh}
	\end{table}

\begin{figure}[ht]
		\includegraphics[scale=.7]{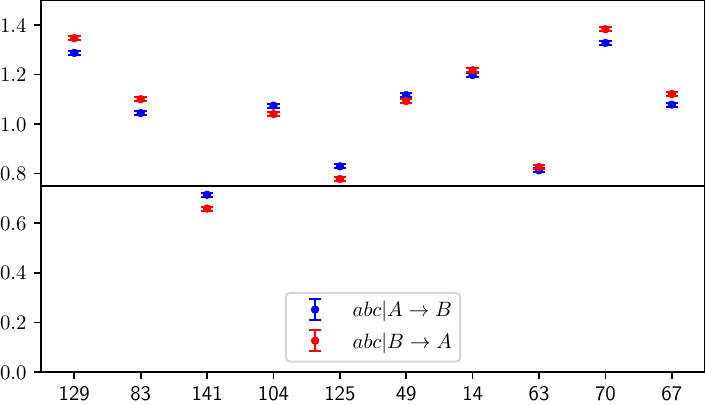}
		\caption{Correlations $\langle abc\rangle/\lambda^2$  on \texttt{ibm\_kingston} with the order $A\to B$ and $B\to A$. The solid line is the perfect
			value $3/4$.
		}
		\label{ABCt}
	\end{figure}

\begin{figure}[ht]
		\includegraphics[scale=.7]{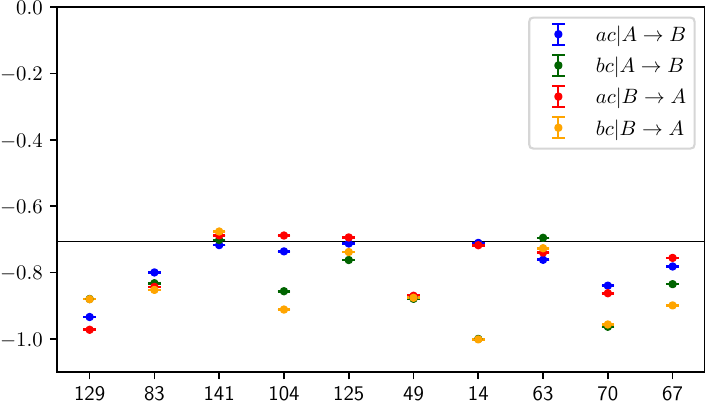}
		\caption{Correlations $\langle ac\rangle/\lambda$  and   $\langle bc\rangle/\lambda$ on \texttt{ibm\_pittsburgh} with the order $A\to B$ and $B\to A$.
			The solid line is the perfect value $-1/\sqrt{2}$.
		}
		\label{ACBCt}
	\end{figure}

\begin{figure}[ht]
		\includegraphics[scale=.7]{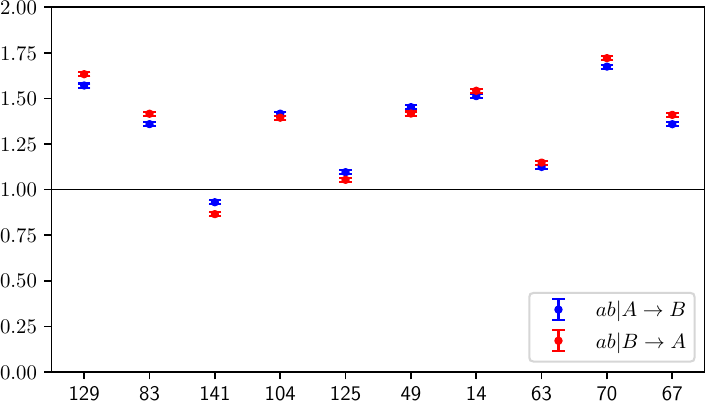}
		\caption{Correlations $\langle ab\rangle/\lambda^2$ on  \texttt{ibm\_kingston}
			with the order $A\to B$ and $B\to A$.
			The solid lines are the perfect value $1$.
		}
		\label{ABt}
	\end{figure}

\begin{figure}[ht]
		\includegraphics[scale=.7]{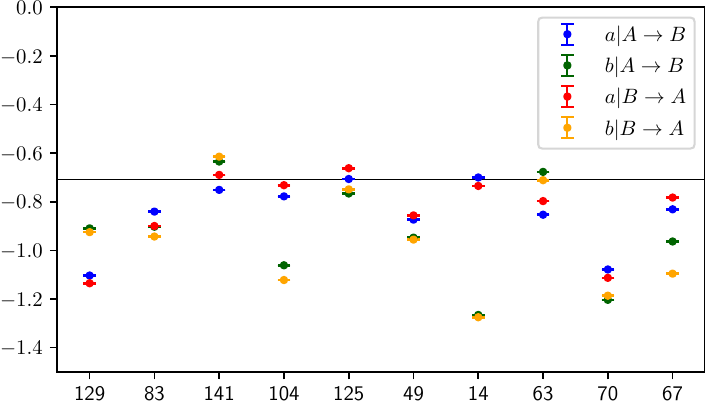}
		\caption{Averages $\langle a\rangle/\lambda$ and $\langle b\rangle/\lambda$ on  \texttt{ibm\_kingston}
			with the order $A\to B$ and $B\to A$.
			The solid lines is the perfect value $-1/\sqrt{2}$.
		}
		\label{At}
	\end{figure}

\begin{figure}[ht]
		\includegraphics[scale=.7]{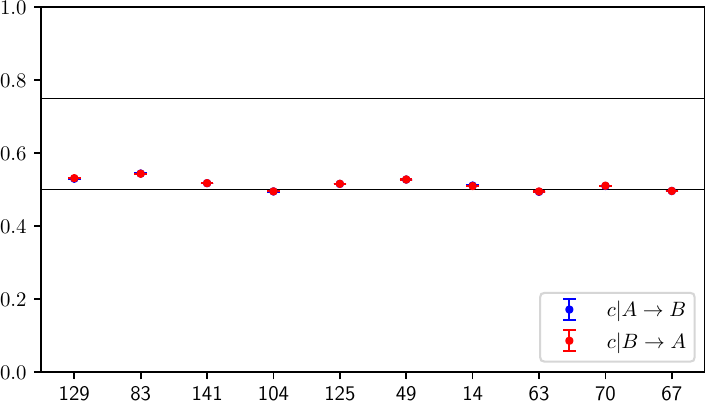}
		\caption{Averages $\langle c\rangle$ on  \texttt{ibm\_kingston} 
			with the order $A\to B$ and $B\to A$.
			The solid line is the perfect value $1/2$ while the replacing weak by projective measurements would change it to $3/4$.
		}
		\label{Ct}
	\end{figure}

The calibration data of the qubits tested on IBM Quantum are presented in 	Table  \ref{fffh}.
For detailed results, all correlations obtained in IBM Quantum experiments, see Figs. \ref{ABCt},\ref{ACBCt},\ref{ABt},\ref{At},\ref{Ct},
and IonQ, Table \ref{tabi}.
	
	 \begin{table}
		\begin{tabular}{*{3}{c}}
			quantity &value&error\\
			\midrule
			(\ref{lgg2}) $A\to B$&1.272 & 0.035\\
			(\ref{lgg2}) $B\to A$&1.289 & 0.035\\
			$\langle abc\rangle_{A\to B}/\lambda^2$&0.829& 0.023\\
			$\langle abc\rangle_{B\to A}/\lambda^2$&0.824& 0.023\\
			$\langle ab\rangle_{A\to B}/\lambda^2$&1.117&0.032\\
			$\langle ab\rangle_{B\to A}/\lambda^2$&1.136&0.032\\
			$\langle ac\rangle_{A\to B}/\lambda$&-0.744&0.002\\
			$\langle ac\rangle_{B\to A}/\lambda$&-0.742&0.002\\
			$\langle bc\rangle_{A\to B}/\lambda$&-0.717&0.002\\
			$\langle bc\rangle_{B\to A}/\lambda$&-0.725&0.002\\
			$\langle a\rangle_{A\to B}/\lambda$&-0.752&0.003\\
			$\langle a\rangle_{B\to A}/\lambda$&-0.744&0.003\\
			$\langle b\rangle_{A\to B}/\lambda$&-0.719&0.003\\
			$\langle b\rangle_{B\to A}/\lambda$&-0.727&0.003\\
			$\langle c\rangle_{A\to B}$&0.507&0.0002\\
			$\langle c\rangle_{B\to A}$&0.507&0.0002\\
			\bottomrule
		\end{tabular}
		\caption{The values and errors from the test on IonQ. Here the first two rows represent the left hand side of Eq. (\ref{lgg2})
			for the order $A\to B$ and $B\to A$. }
		\label{tabi}
	\end{table}

\end{document}